\documentclass[aps,prl,twocolumn,superscriptaddress,showpacs]{revtex4}
\usepackage{graphicx}  

\begin{document}

\title{Controllable spin-current blockade in a Hubbard chain}

\author{Yao Yao}\affiliation{Surface Physics Laboratory (National Key Laboratory)
and Department of Physics, Fudan University, Shanghai 200433,
China}
\author{Hui Zhao}\affiliation{Surface Physics Laboratory (National Key Laboratory)
and Department of Physics, Fudan University, Shanghai 200433,
China}
\author{Joel E. Moore}\affiliation{Department of Physics, University of California, Berkeley, California 94720,
USA}\affiliation{Materials Sciences Division, Lawrence Berkeley
National Laboratory, Berkeley, California 94720, USA}
\author{Chang-Qin Wu}
\email[Email: ] {cqw@fudan.edu.cn} \affiliation{Surface Physics
Laboratory (National Key Laboratory) and Department of Physics,
Fudan University, Shanghai 200433, China}

\date{\today}
\begin{abstract}
We investigate the spin/charge transport in a one-dimensional
strongly correlated system by using the adaptive time-dependent
density-matrix renormalization group method. The model we consider
is a non-half-filled Hubbard chain with a bond of controllable
spin-dependent electron hoppings, which is found to cause a
blockade of spin current with little influence on charge current.
We have considered (1) the spread of a wave packet of both spin
and charge in the Hubbard chain and (2) the spin and charge
currents induced by a spin-dependent voltage bias that is applied
to the ideal leads attached at the ends of this Hubbard chain. It
is found that the spin-charge separation plays a crucial role in
the spin-current blockade, and one may utilize this phenomenon to
observe the spin-charge separation directly.

\end{abstract}


\maketitle

In one-dimensional (1d) strongly correlated systems, an essential
phenomenon is the spin-charge separation (SCS),~\cite{Luttinger}
which is believed to play a central role in 1d transport~\cite{Anderson}.  To study transport problems in 1d systems, the characteristics of SCS must be taken into account.
However, due to the limitations of existing methods, the discussion in the
past has been limited to a few simple cases; an example is the significant work
by Kane and Fisher\cite{Kane} on scaling properties of tunneling through a spin-symmetric point
impurity in a fermion system. Hence more
powerful methods are needed to compute transport properties beyond scaling and to treat a general interacting Hamiltonian.  Furthermore, spin-dependent
transport problems have attracted increasing interest in the last
two decades: many proposed spintronics devices~\cite{Giant,SpinPolarized} require the manipulation
of spin currents, which are naturally decoupled from charge currents in 1d systems. On the other hand, a number of experimental works have sought to observe the phenomenon.\cite{SCSExp} Rapid progress in ultracold atomic gas experiments\cite{OL} makes it
possible to see SCS in a new context.\cite{OL_SCS}

Recently, the adaptive time-dependent density-matrix renormalization
group (t-DMRG)\cite{adaptive1,adaptive2} was developed by combining
the DMRG method\cite{white} with quantum information concepts. The
key idea of this method is to break up the evolution operator with
Trotter decomposition,\cite{Trotter} then apply it to
the states within a DMRG configuration.  At the same time, other
real-time evolution methods within DMRG\cite{static,xiang,Peter,TST}
were also proposed. By use of these numerical methods, there have
been investigations on transport properties in 1d
strongly-correlated or impurity systems, including spin-$1/2$
chains,\cite{SpinChain} Bose-Hubbard model,\cite{boson} and
conductance analysis.\cite{conductance} The dynamical problems with
impurities were also studied widely using static DMRG method
embedding with persistent current\cite{embedding,persist} and
functional renormalization group.\cite{fRG,persist}  An interesting result that partly motivates our study was the study of SCS by Kollath, Schollwoeck, and Zwerger.\cite{SCS} In conventional treatment with the bosonization method, only low energy excitations were considered. The above study goes beyond the low energy excitation spectrum by considering the evolution of a "big" (multiparticle) wave packet that shows the SCS phenomenon.\cite{SCS}

In this Letter, we propose to consider a non-half-filled Hubbard
chain in which one special bond has controllable spin-dependent
electronic hoppings, motivated by the development of optical lattices
of ultracold atoms in which all hoppings can be controlled. By using the adaptive t-DMRG method,
we simulate the spread of a wave packet as well as the spin and
charge currents under a spin-dependent voltage bias. We find that
the spin-current blockade can be realized by adjusting the
spin-dependent hopping on that special bond while the charge
current has not been affected. A possible application of this
phenomenon is discussed.

The system we are considering is described by the following
Hamiltonian,
\begin{equation}
H_S=-\sum_{i,\sigma}t_{i,i+1}^\sigma(c^\dagger_{i,\sigma}c_{i+1,\sigma}+{\rm
h.c.})+U\sum_i n_{i,\uparrow}n_{i,\downarrow},
\end{equation}
where $c^\dagger_{i,\sigma} (c_{i,\sigma})$ creates (annihilates)
an electron with spin $\sigma (=\uparrow,\downarrow)$ on the
$i$-th site, $n_{i,\sigma}(\equiv c^\dagger_{i,\sigma}
c_{i,\sigma})$ is the corresponding electron number operator, the
electron hopping constants $t_{i,i+1}^\sigma\equiv t_0$ on all bonds
but a special bond ($i=l_s$), where $t_{i,i+1}^\sigma(=t_\sigma)$
is an adjustable spin-dependent quantity, which introduces a local
magnetic moment, like a spin-dependent Anderson impurity. Without
the special bond, the Hamiltonian is nothing but the usual Hubbard
model with $t$ being the electron hopping constant between nearest
neighbor sites and $U (>0)$ the on-site repulsive Coulomb
interaction. Without loss of generality, we will always keep
$t_\uparrow=t_0$ while $t_\downarrow$ is adjusted from $t_0$
to 0.
We present a proposal in Fig.~1 on the realization of a Hubbard
chain with a special bond of controllable spin-dependent
electronic hopping in an optical lattice. In the optical lattice,
the spin-up and -down atoms could be trapped in a pair of parallel
period potential wells created by interfering laser
beams.\cite{OL} Hence, one might change the laser beam to rotate
half of spin-down atoms around the spin-up atom axis by an angle
$\theta$, which reduces the spin-down atom hopping at the special
bond. It is clear that $t_\downarrow$ of this special bond in
Eq.~(1) could be easily adjusted by changing $\theta$.

\begin{figure}
\includegraphics[angle=0,scale=0.35]{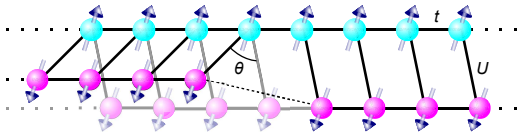}
\caption{The realization of a Hubbard chain with a special bond of
controllable spin-dependent electronic hopping in an optical
lattice. The rotation of the left spin-down atoms around the
spin-up atom axis reduces the spin-down atom hopping at the
special bond, {\it i.e.}, the $t_\downarrow$ in Eq.~(1) decreases
as the angle $\theta$ increases. \label{fig1}}
\end{figure}

First, we consider the spread of a wave packet as
done by Kollath et al.~\cite{SCS}, but in a Hubbard chain with
a bond of controllable spin-dependent electron hoppings, described
by Eq.(1). The wave packet is induced by the following
spin-dependent Gaussian potential
\begin{equation}
H_P=-P\sum_{i} {\exp\left[-\frac{(i-l_p)^2}{2{l_d}^2}\right]
\hat{n}_{i,\uparrow}},
\end{equation}
which acts only on the spin up electrons, so that it carries both
of spin and charge. Then the potential will be switched off and
the wave packet will be spread according to the time-dependent
many-body Schr\"{o}dinger equation.  Clearly $P$ determines the
potential strength, $l_p$ the center of the induced wave packet and
$l_d$ its width. A small value of $l_d$ (\emph{e.g.}, $\le$1)
corresponds to a local potential at a single site as suggested by
Anderson,\cite{Anderson} the case we will consider here.

  In the following, we apply the t-DMRG method with second order
Trotter decomposition to simulate the dynamical evolution of this
wave packet. The number of sites ($L$) is taken to be $176$
while the number of electrons ($N_e$) is $116$; the corresponding
filling factor ($N_e/(2 L)$) is about 1/3. The time step is taken as
$0.04$ (in unit of $\hbar/t_0$) and the number of kept DMRG states
($M$) is chosen to be large enough to ensure the error being less
than $O(10^{-3})$.
The spin and
charge currents, the quantities we focus on in this
calculation, are defined as
$J_s(j)=J_{\uparrow}(j)-J_{\downarrow}(j)$ and
$J_c(j)=J_{\uparrow}(j)+J_{\downarrow}(j)$, separately, where
\begin{eqnarray}
J_{\sigma}(j)\equiv it_{j,j+1}^\sigma\langle
c^\dagger_{j,\sigma}c_{j+1,\sigma}
-c^\dagger_{j+1,\sigma}c_{j,\sigma}\rangle.
\end{eqnarray}

\begin{figure}
\includegraphics[angle=0,scale=0.8]{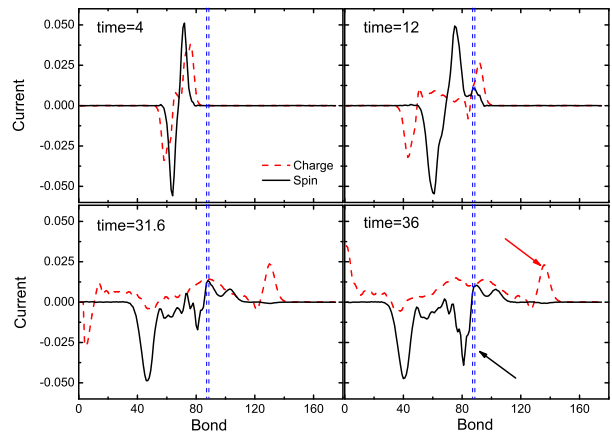}
\caption{The spin (solid line) and charge (dashed line) current in
the spread of a wave packet in a Hubbard chain with a special bond
with $U=8t_0, l_s=83$ and $P=0.8t_0, l_d=1, l_p=62$. The location of
this special bond is indicated by a double dashed (blue) line.
Positive values imply the current flows to the right while negative
values the current flows to the left. The dashed (red) arrow
indicates the charge transmitted through the special bond while the
solid (black) one the spin reflected from the special bond.
\label{fig2}}
\end{figure}
  In Fig.~\ref{fig2}, we show the spin and charge current
at various times for $t_\downarrow=0$ . The initial potential
locally on the spin up generates simultaneously spin and charge wave
packet, and then the wave packets will split into two parts and
propagate to opposite directions respectively.\cite{SCS} Since the
potential strength $P$ we took is not large, the spin and charge
density of this wave packet is very small, but the currents defined
in Eq.(3) show the propagation of this wave packet very clear. In
Fig.~\ref{fig2}(a), we show the current distribution at a time
before the spin and charge reach the special bond.
The split of the two peaks indicates the different speeds of spin and charge excitations, so that the
spin-charge separation is observed clearly. Next we'll see clearly
from the figure, the charge current goes through the special bond
almost freely, while the spin current is blocked by the special
bond. A spin current reflecting at the special bond is shown in the
figure by its value being changed from positive to negative. The
change of charge currents from negative to positive indicates the
charge reflection at the left end since we use an open boundary
condition for the chain.

We argue that, in the following, the spin-current blockade we
observe here only happens in a strongly correlated system. It is
different from the ``spin blockade'' effect,\cite{SpinBlockade}
which is merely spin-related Coulomb blockade. To understand the
phenomenon we observed above, we consider the large $U$ limit of
the Hubbard-like model in Eq.~(1), which leads to the so-called
$t-J$ model $H_{t-J}=H_t+H_J$, where $H_t$ is the hopping term in
Eq.~(1) and
\begin{equation}
H_J=-\frac{1}{U}\sum_{ijkss'}t_{ij}^st_{jk}^{s'}c_{is}^\dagger
c_{js}n_{j\uparrow}n_{j\downarrow}c_{js'}^\dagger c_{ks'}
\end{equation}
with $t_{ij}^s\equiv
t_{i,i+1}^s\delta_{i,j-1}+t_{i-1,i}^s\delta_{i,j+1}$.\cite{auerbach}
This model works on the space that has projected out all
configurations with at least one doubly occupied site for a
less-than-half filling system and is responsible for the
low-energy excitations of the model in Eq.~(1). It is well known
that the hopping term $H_t$ is responsible for the charge
excitations while $H_J$ controls the spin excitations, and that
for the usual Hubbard model, $H_J$ corresponds to a Heisenberg spin chain.
Writing
\begin{equation}
\vec{S}_i\cdot\vec{S}_j \equiv S_i^z S_j^z+\frac{1}{2}(S_i^+
S_j^-+S_i^- S_j^+),
\end{equation}
the second term is responsible for the spin-exchange process that is
necessary for an $S_z$ spin current. Based on $H_J$ in Eq.(4) from the
large-$U$ expansion, we show the spin-exchange process in Fig.~3, in
which there is a virtual intermediate state,
so electronic hoppings for both spins are necessary for the
process via a virtual state. Then it is clear that spin current is
blocked at the special bond when $t_\downarrow=0$ since only spin-up
electrons can hop across the special bond.
\begin{figure}
\includegraphics[angle=0,scale=0.35]{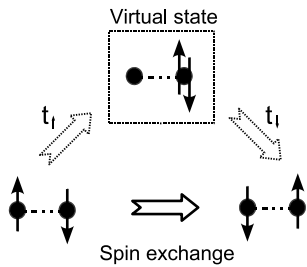}
\caption{A spin-exchange process that is necessary for the spin
current in the strong interaction limit.\label{fig3}}
\end{figure}

Now, we come to the calculation of spin and charge currents under a
spin-dependent voltage bias, for which we consider the system of a
Hubbard chain with a special bond is attached with two ideal leads at
its two end, that is described by the following Hamiltonian,
\begin{equation}
H=H_S+H_{lead}+H_{int},
\end{equation}
where $H_S$ has been given in Eq.(1) and the summation over site index runs
from 1 to $L$ with $L$ being the size of the Hubbard chain,
$H_{lead}\equiv H_{L}+H_{R}$ and
\begin{equation}
H_\alpha=-t_0\sum_{i,\sigma \in \alpha}(c^\dagger_{i,\sigma}c_{i\pm 1,\sigma}+{\rm
h.c.})-\sum_{i,\sigma \in \alpha}V_{\sigma}^\alpha n_{i,\sigma},
\end{equation}
with $\alpha=L$ or $R$ and the sign $\pm$ taking $- (+)$ for $\alpha=L~(R)$,
and
\begin{equation}
H_{int}=-t_0\sum_\sigma (c^\dagger_{0,\sigma}c_{1,\sigma}+c^\dagger_{L,\sigma}c_{L+1,\sigma}+{\rm
h.c.}).
\end{equation}
The spin-dependent voltage bias $V_\sigma\equiv
V_\sigma^L-V_\sigma^R$ applied at the two leads turns on at $t=0$,
so that currents appear gradually at the same time and reach constant values finally. In the calculation, we take $V_\downarrow=0$ and
$V_\uparrow=0.1t_0$, without loss of generality, to induce both
spin and charge currents through the Hubbard chain.

\begin{figure}
\includegraphics[angle=0,scale=0.5]{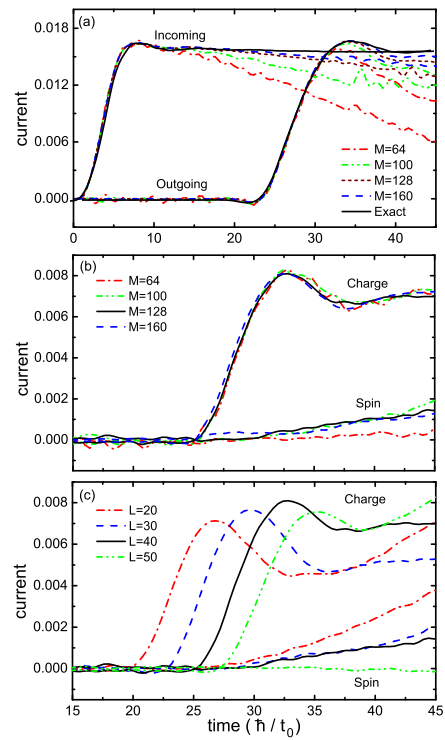}
\caption{The spin and charge currents at various times for different
numbers of kept DMRG states ($M$) or Hubbard chain lengths ($L$).
(a) Both incoming and outgoing spin/charge currents are shown for
$U=0$ and $L=40$. (b) Only the outgoing currents are shown for
$U=8t_0$ and $L=40$. (c) The outgoing currents for different chain
lengths $L$, and $U=8t_0$ and $M=160$.\label{fig4}}
\end{figure}

Now we show in Figure~4 the spin and charge currents at various
times for different Hubbard chain lengths ($L$) or the numbers of
kept DMRG states ($M$) in the calculations induced by the
spin-dependent voltage bias. First the Hubbard interaction is
switched off for the test of the calculation precision for a chain
of $L=40$ without the special bond. The lengths of ideal leads
attached the chain are taken to be long enough so that the results
are sufficiently insensitive to it. The incoming and outgoing
currents are defined as the corresponding currents at the left and
right interfaces according to Eq.~(3), respectively. At $U=0$, the
spin and charge currents are the same since no interaction between
electrons of different spins exists. Furthermore, the currents can
be calculated exactly at the case, which is shown as solid lines in
(a). A time lag between the incoming and outgoing currents is
observed due to the spin/charge transportation from the left chain
end to the right one. Finally, steady transport is reached at
$t=40 (\hbar/t_0)$. In Fig.~4(a), we also show the results obtained
from the t-DMRG method by keeping various numbers of DMRG states. It
can be seen that the curves obtained when the number of kept DMRG states ($M$)
is 128 or more are very close to the exact one (solid
lines), while the convergence for the outgoing current is faster than
that of the incoming one, the reason being that the voltage bias
is applied only at the left interface for convenience in this
calculation.

Next, in Fig.~4(b) we show both spin and charge outgoing currents
for a Hubbard chain of $L=40$ and $U=8t_0$ with the special bond for
a number of different kept DMRG states ($M$). It is clearly seen
that the convergence for a Hubbard chain is much better than that of
the $U=0$ chain with increasing the number of kept DMRG states
($M$). In a sufficiently long time, it can be seen that while the
charge current passes through the chain the spin current is blocked
and only a little of it passes. This result is consistent with the
spread of a wave packet studied above. In Fig.~4(c) we give the
currents for different lengths ($L$). We observe that a Hubbard
chain of moderate length $L~(=30 - 40)$ is optimal, for a shorter
chain is accompanied by a serious finite-size effect and a longer chain by a
large accumulated error since it takes more time for spin/charge
to travel from one end to the other.

Finally, we give the dependence of spin/charge transport through
the Hubbard chain on the parameters $t_\downarrow$ and $U$ in
Fig.~5. The transmission rates are defined as the ratio between
transmitted and incident currents, where the former is taken from
the average value of the outgoing current at a period when the
current is almost steady and the latter is the same of the incoming
current at the period before the current reaches the right end.
\begin{figure}
\includegraphics[angle=0,scale=0.45]{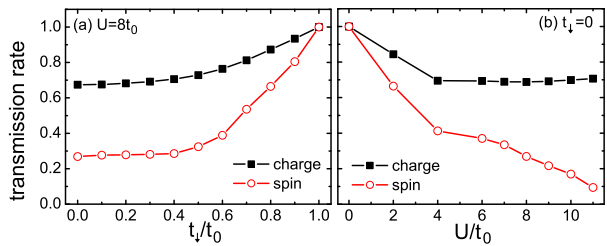}
\caption{The dependence of transmission rates (see text for
definition) on (a) $t_\downarrow$ ($U=8t_0$) and (b) $U$
($t_\downarrow=0$). \label{fig5}}
\end{figure}
From Fig.~\ref{fig5}(a), it can be seen that the transmission rate
changes little for $t_\downarrow<0.5$. On the other hand, Fig.~\ref{fig5}(b) shows that the on-site Hubbard repulsion $U$ plays a crucial role in the observed spin-current blockade. When $U$
increases, the charge transmission rate changes little in the strongly-correlated
regime, namely, $U>4t_0$, and the spin rate decreases very quickly
and reaches about 0.1 for $U>11t_0$. It implies the spin current is
nearly completely blocked by the special bond.

An experimental observation of spin-charge separation in cold
Fermi gases has been proposed by Kollath, et al.,\cite{SCS} in
which the energy scale $k_B T$ should be much smaller than the
Mott energy gap to ensure that thermal activation does not destroy
the Mott-insulating behavior. But this is not required in our case
since the Mott gap is zero. Furthermore, an important advantage is
that the spin-current blockade could be realized by adjusting only
one parameter ($t_\downarrow$), which simplifies experimental
observation.

In summary, we have investigated the spin and charge transport in
a Hubbard chain with a bond of controllable electronic hopping. We
find the spin current can be blocked by this special bond while
charge current passes through the bond freely. It is found that a
large Hubbard $U$ is required for the observed blockade since the
spin-charge separation plays a crucial role in it.

This work was supported by the NSF of China, the MST of China
(2006CB921302), MOE of China (B06011), the Western Institute of
Nanoelectronics, and the EC Project OFSPIN (NMP3-CT-2006-033370).
C.Q.W. is grateful to UC Berkeley for the hospitality during his
visit there.

\end{document}